\documentclass[
preprint,
amsmath,
amssymb,
]{revtex4-2}

\usepackage{graphicx}
\usepackage{dcolumn}
\usepackage{bm}
\usepackage{CJKutf8}
\usepackage[dvipsnames]{xcolor}

\begin{document}

\title{Assessing the electronic excitation spectra of chromium, palladium and samarium from their stopping quantities }

\author{Fan Cheng [\begin{CJK*}{UTF8}{gbsn} 
程凡
\end{CJK*} ]$^{1,2,3}$}
\author{Pablo de Vera$^1$}%
 \email{pablo.vera@um.es}
\author{Rafael Garcia-Molina$^1$}
\affiliation{%
 $^1$Departamento de Física -- Centro de Investigación en Óptica y Nanofísica, Universidad de Murcia, E-30100 Murcia, Spain
}%
\affiliation{%
 $^2$Key Laboratory of Materials Physics, Institute of Solid State Physics, HFIPS, Chinese Academy of Sciences, 230031 Hefei, People’s Republic of China
}%
\affiliation{%
 $^3$University of Science and Technology of China, 230026 Hefei, People’s Republic of China
}%

\date{\today}

\begin{abstract}
The electronic excitation spectrum of a material characterises the response to external electromagnetic perturbations through its energy loss function (ELF), which is obtained from several experimental sources that usually do not completely agree among them. In this work, we assess the available ELF of three metals, namely chromium, palladium, and samarium, by using the dielectric formalism to calculate relevant stopping quantities, such as the stopping cross sections for protons and alpha particles, as well as the corresponding electron inelastic mean free paths. The comparison of these quantities (as calculated from different sets of ELF) with the available experimental data for each of the analyzed metals highlights the promising capability of the recently proposed reverse Monte Carlo method for the determination of the ELF. This work also analyzes the contribution of different electronic shells to the electronic excitation spectra of these materials, and reveals the important role that the excitation of ``semi-core'' bands plays on the energy loss mechanism for these metals.  
\end{abstract}

\maketitle


\section{Introduction}
\label{sect:introduction}
The response of a material to electromagnetic perturbations, induced either by an external radiation or by a field, depends on its electronic excitation spectrum, which is encoded in the so-called energy loss function (ELF), ${\rm Im}[-1/\varepsilon]$, being $\varepsilon$ the complex dielectric function of the material \cite{Nikjoo2012}. A proper knowledge of the ELF for a given material is of paramount importance for either applied or basic purposes \cite{Sigmund2006,Sigmund2014}. 


Even though theoretical calculations of the ELF structure (i.e., its energy and momentum dependence) are nowadays feasible \cite{Pitarke2000,Botti2007,Yan2011,Taioli2021JPCL,Pedrielli2021,Taioli2023}, they require a considerable computational effort, so this quantity is mostly obtained experimentally from optical measurements (such as refractive index, extinction coefficient, reflectivity...) or by electron spectroscopy.
However, the resulting data for the same material do not always coincide (within an admissible range), which can lead to substantial differences in the quantities that depend on the ELF.

In this work we use the dielectric formalism \cite{Fermi1940,Lindhard1954,Ritchie1959} to calculate well defined stopping quantities (stopping cross sections for protons and alpha particles, as well as electron inelastic mean free paths) for chromium, palladium and samarium, using as inputs their ELF as provided from various sources, which rely on different experimental techniques. These materials, apart from presenting excitation spectra characteristic of those of the heavy metals (showing a prominent contribution in the spectrum coming from ``semi-core'' levels), are found in numerous technological applications. Chromium is commonly used for the production of a wide variety of stainless steels. Both chromium and palladium are useful catalysts for chemical reactions. Palladium is also found in the electronics industry, as a component of ceramic capacitors. Samarium finds applications in magnetic devices as well as a neutron absorber in nuclear reactors. Moreover, for these metals there are several discrepant sources for their ELF: optical data \cite{Palik1999}, transmission electron energy loss spectroscopy \cite{Vehenkel1974}, reflection electron energy loss spectroscopy \cite{Werner2009,Tahir2014}, as well as the latter analyzed by means of the recent reverse Monte Carlo method \cite{Xu2018,Yang2023}. The comparison of these calculated stopping quantities with available experimental data is used to asses which are considered the most reliable ELF for each metal discussed in this work.

This paper is organized as follows. Section \ref{sect:Theory} contains the basic ingredients of the theoretical framework used to calculate the ion stopping cross sections and electron inelastic mean free paths. In Section \ref{sect:Results} we compare with available experimental data our calculated stopping quantities derived from different ELF sources. Finally, in Section \ref{sect:Conclusions} we draw the main conclusions regarding the different experimentally determined electronic excitation spectra of Cr, Pd, and Sm.

\section{Theoretical framework}
\label{sect:Theory}

The dielectric formalism \cite{Fermi1940,Lindhard1954,Ritchie1959} constitutes a widely used theory to calculate the stopping quantities characterizing how swift energetic charged particles interact with a material. When the velocity of the particles becomes comparable to that of the target electrons, the first Born approximation in which the dielectric formalism is based breaks, and other methodologies to calculate quantities such as the stopping power are more appropriate, like real-time time dependent density functional theory \cite{Yost2017,Kononov2023,Matias2024,Matias2024b}. 

In this work we are interested in the inelastic mean free path $\lambda (T)$ for an electron with mass $m$ and energy $T$ which, within the dielectric formalism, is calculated as:
\begin{equation}
    \lambda^{-1} (T) = \frac{e^2}{\pi \hbar^2} \frac{m}{T} \int_{E_-}^{ E_+} {\rm d}E \int_{k_{-}}^{k_{+}} \frac{{\rm d}k}{k} \, {\rm Im}\left[ \frac{-1}{\varepsilon(k,E)} \right] \, \mbox{,}
    \label{eq:IMFP}
\end{equation}
and the stopping power $S_{\rm q} (T)$ (mean energy loss per unit path length) for an ion with mass $M$, atomic number $Z$, charge state $q$, and energy $T$:
\begin{equation}
    S_q (T) = \frac{e^2}{\pi \hbar^2} \frac{M}{T} \int_{E_-}^{E_+}E\, {\rm d}E \int_{k_{-}}^{k_{+}}  \frac{{\rm d}k}{k} [Z-\rho_q(k)]^2 \, {\rm Im}\left[ \frac{-1}{\varepsilon(k,E)} \right] \, \mbox{,}
    \label{eq:Sq}
\end{equation}
where $\rho_q(k)$ is the Fourier transform of the electronic density corresponding to a charge state $q$ and $\varepsilon(k,E)$ is the target dielectric function. The values of the integration limits, resulting from conservation laws, are provided and discussed in \cite{deVera2023}.

The electronic cloud around an ion nucleus is deformed due to the electric field induced in the target at the position of the projectile \cite{HerediaAvalos2002a}, therefore Eq. \eqref{eq:Sq} must be modified to account for this effect, resulting in $S_{{\rm pol},q} (T)$ as detailed in \cite{deVera2023}.

Due to electron capture and loss by the projectile as it moves through the target, an additional contribution $S_{\rm {C\& L}} (T)$ must be considered. Taking into account all these factors, the total electronic stopping power can be written as \cite{deVera2023}
\begin{equation}
S(T) = \sum_q \phi_q(T) S_{{\rm pol},q}(T) + S_{\rm C\&L} \, \mbox{,}
\label{eq:Sions}
\end{equation}
where $\phi_q(T)$ is the charge fraction of projectiles with charge state $q$.

Leaving aside the technical details of the calculation of $\lambda(T)$ and $S_{\rm q} (T)$, already discussed in  \cite{deVera2023}, what is relevant for our present work is that these stopping quantities depend on the target electronic excitation spectrum through its energy loss function (ELF), ${\rm Im}\left[-1/\varepsilon(k,E) \right]$, which depends on the momentum $\hbar k$ and energy transfers $E=\hbar \omega$ of the target electronic excitation spectrum.

In order to evaluate the integrals that appear in the calculations of the stopping quantities, it is convenient to have an analytical expression of the ELF for all values of $k$ and $E$.  This can be achieved by employing the MELF-GOS (Mermin energy loss function - generalized oscillator strengths) methodology, described in detail elsewhere \cite{HerediaAvalos2005a, deVera2023}, which provides the ELF of a material for all values of $(k, E)$ from available experimental ELF in the optical limit (i.e., at $k = 0$). Within the MELF-GOS methodology, the outermost features of the optical ELF are fitted by means of Mermin-type energy loss functions (MELF), each having a characteristic position $\hbar \omega_i$, width $\hbar \gamma_i$ and intensity $A_i$. The higher energy part of the ELF, corresponding to the excitation of the atomic-like inner shells, is reproduced by means of hydrogenic generalized oscillator strengths (GOS).

In the present work, we have applied the MELF-GOS methodology to obtain the ELF of chromium, palladium, and samarium from different sets of available optical ELF for each material. 
Table \ref{tab:ELF} shows the sources of the ELF of the three metals discussed in this work, which are the following: 
optical data (OD) from Palik and Ghosh's compilation \cite{Palik1999}, transmission electron energy loss spectroscopy (TEELS) by Wehenkel and Gauth\'e \cite{Vehenkel1974},
reflection electron energy-loss spectroscopy (REELS) by Werner \textit{et al.} \cite{Werner2009} and Tahir \textit{et al.} \cite{Tahir2014}, and REELS combined with the reverse Monte Carlo technique (RMC) by Xu \textit{et al.} \cite{Xu2018} and Yang \textit{et al.} \cite{Yang2023}. 


\begin{table}  
    \centering  
    \renewcommand{\arraystretch}{1.2}
    \setlength{\tabcolsep}{10pt}
    \begin{tabular}{|c|c|c|}
        \hline  
        Cr & Pd & Sm  \\  
        \hline  \hline
        Palik and Ghosh \cite{Palik1999} (OD) & Palik and Ghosh \cite{Palik1999} (OD) & Yang \textit{et al.} \cite{Yang2023} (RMC) \\  
        Xu \textit{et al.} \cite{Xu2018} (RMC) & Xu \textit{et al.} \cite{Xu2018} (RMC) &  \\ 
        Wehenkel and Gauth\'e \cite{Vehenkel1974} (TEELS) & Werner \textit{et al.} \cite{Werner2009} (REELS) & \\
            & Tahir \textit{et al} \cite{Tahir2014} (REELS) & \\    
        \hline  
    \end{tabular}%
    \caption{Sources of the ELF for each metal. Acronyms refer to the experimental methodology employed. OD: optical data; RMC: reverse Monte Carlo; TEELS: transmission electron energy loss spectroscopy; REELS: reflection electron energy loss spectrocopy.}
    \label{tab:ELF}
\end{table}

\begin{figure}
	\centering
	\includegraphics[width=.55\textwidth]{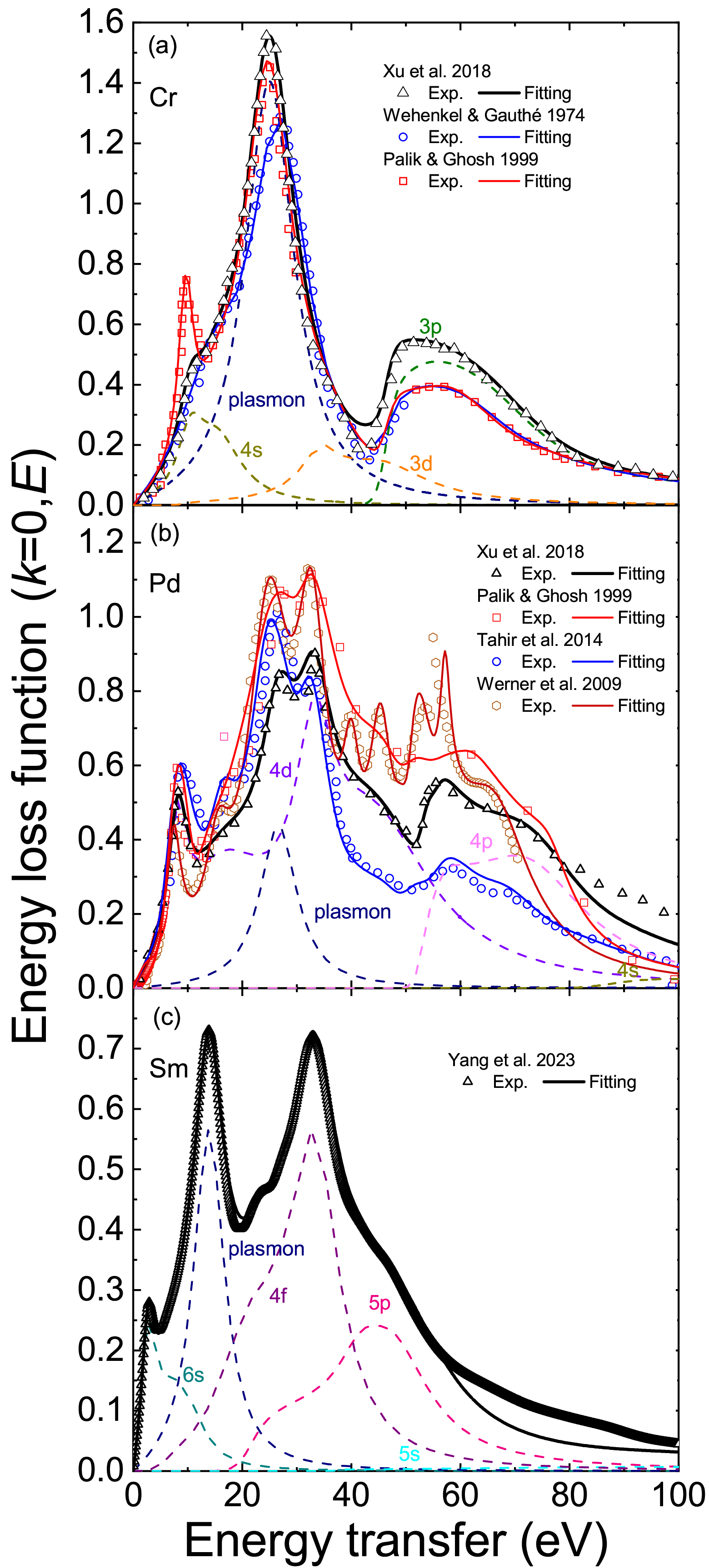}
	\caption{Optical energy-loss function, ELF ($k = 0,E$), of (a) Cr, (b) Pd, and (c) Sm, as a function of the energy transfer $E$. Symbols depict experimental data, while solid lines represent the MELF-GOS fit to each set of experimental data. The dashed lines represent different contributions to the ELF of Cr and Pd from Xu \textit{et al.} \cite{Xu2018} and to the ELF of Sm from Yang \textit{et al.} \cite{Yang2023}.}
	\label{FIG:1}
\end{figure}

\section{Results and discussion}
\label{sect:Results} 

Figure \ref{FIG:1} gathers all the experimental sources for the optical ELF of (a) Cr, (b) Pd, and (c) Sm, shown by different symbols. In general terms, all experimental optical ELF of Cr and Pd present excitation peaks at similar energies for each metal (for Sm, only one set of ELF data is available). However, the widths and intensities of the peaks in the ELF of each metal can be considerably different, what will influence the calculated energy loss quantities. As these data are available up to some limiting energy (typically below $100$ eV), the ELF is extended to larger energies by means of the atomic form factors provided by the NIST's FFAST database \cite{Chantler2005}.

The consistency of an ELF, as well as the nature of its different excitations, can be assessed with the help of sum rules \cite{Smith1978,deVera2023}, such as the Kramers-Kronig (KK): 
\begin{equation}\frac{2}{\pi} \int_0^{\infty} \textrm{d}E' \, \frac{1}{E'} \, \textrm{Im}\left[ \frac{-1}{\varepsilon(k=0,E')} \right] = 1 .
\label{eq:KK}
\end{equation}
\noindent and the $f$-sum rule:
\begin{equation}
N_{\rm eff}(E) = \frac{m}{2\pi^2 \hbar^2 e^2 \mathcal{N}} \int_{0}^{E} \textrm{d}E' \, E' \, \textrm{Im}\left[ \frac{-1}{\varepsilon(k,E')} \right] \mbox{ , }
\label{eq:f-sum_ELF}
\end{equation}
\noindent where $\mathcal{N}$ is the atomic density of the target. In the former, the integral should converge to a number close to $1$. In the latter, $N_{\rm eff}(E)$ denotes the effective number of electrons participating in excitations with energy transfers smaller than or equal to $E$, which should converge to the atomic number of the target material in the limit $E \rightarrow \infty$. As suggested in Refs. \cite{deVera2019JPCC,deVera2023}, partial $f$-sum rules can be used to determine the effective number of electrons involved in each feature of the ELF, to estimate the nature of the excitation, by performing the integral of Eq. (\ref{eq:f-sum_ELF}) over the particular MELF contributing to this excitation.

Another important energy loss quantity can be calculated by integration of the ELF, namely the mean excitation energy $I$, entering the Bethe equation for the stopping power \cite{Inokuti1971,Fano1963}:
\begin{equation}
\ln I = \frac{\displaystyle \int_{0}^{\infty} {\rm d}E' \, E' \ln E' \, {\rm Im}\left[{\frac{-1}{\varepsilon(k=0,E')}}\right]}
{\displaystyle \int_{0}^{\infty} {\rm d}E' \, E'  \, {\rm Im}\left[\frac{-1}{\varepsilon(k=0,E')}\right]} \, \mbox{.}
\label{eq:I}
\end{equation}
This quantity will be calculated for the three metals, and compared to the values reported in the literature \cite{ICRU37}.

In the following sections, the different ELF available for each of the studied metals will be analyzed and assessed, comparing the stopping powers for protons and helium ions and the inelastic mean free paths for electrons derived from them with the available values got from experiments.

\subsection{Chromium}
\label{sect:Cr} 

The optical ELF of Cr is relatively clean (Fig. \ref{FIG:1}(a)), with all experimental sources providing a very intense peak around $25$ eV, slightly less intense in the case of Wehenkel and Gauthé's TEELS experiments \cite{Vehenkel1974}. While the OD by Palik and Ghosh \cite{Palik1999} show a narrow peak around $10$ eV, the results by Wehenkel and Gauthé \cite{Vehenkel1974} and by Xu \textit{et al.}'s RMC \cite{Xu2018} just give a shoulder. A broad feature with an onset around $40$ eV is given by all three datasets, but with Xu \textit{et al.}'s data displaying it rather more intense.

One or several MELF have been assigned to the excitation of electrons from a particular outer band of the metal (see Table \ref{tab:comparison-Cr}), and one of them is identified as the plasmon resonance by comparing its position to the plasmon energies collected by Egerton \cite{Egerton2011}. The different bands are guessed by means of the growing values of their binding energies \cite{Chantler2005} (also indicated in Table \ref{tab:comparison-Cr}). The excitations from the $2p$ band and inner ones (hereafter referrer as inner shells) are modelled by means of the hydrogenic GOS. The different contributions to the outer-shell ELF are depicted by dashed lines in Fig. \ref{FIG:1}(a).

\begin{table}  
    \centering  
    \renewcommand{\arraystretch}{1.5}
    \setlength{\tabcolsep}{10pt}
    \resizebox{\textwidth}{!}{%
    \begin{tabular}{|c|c|c|c|c|c|}
        \hline  
        Outer shell & Expected electrons & Xu \textit{et al.} \cite{Xu2018} & Palik and Ghosh \cite{Palik1999} & Wehenkel and Gauthé \cite{Vehenkel1974} & {Binding energy (eV)} \\  
        \hline  \hline
        $4s$ & 1 & 0.42 & 0.41 & 0.50 & {-} \\  
        \hline
        Plasmon & {-} & 3.38 & 3.35 & 3.74 & {-} \\ 
        \hline
        $3d$ & 5 & 1.03 & 0.45 & 0.26 & 8.71 \\
        \hline  
        $3p$ & 6 & 8.23 & 6.44 & 6.67 & 46.26 \\  
        \hline
        $3s$ & 2 & 1.66 & 3.31 & 2.54 & 74.01 \\  
        \hline \hline
        Inner shell &  &  &  &  &  \\  
        \hline \hline
        $2p$ & 6 & 6.51 & 6.51 & 6.51 & 579.03 \\  
        \hline  
        $2s$ & 2 & 1.57 & 1.57 & 1.57 & 687.60 \\  
        \hline
        $1s$ & 2 & 1.34 & 1.34 & 1.34 & 5988.92 \\  
        \hline \hline 
        Total &  24 & 24.15 & 23.64 & 23.19 & {-} \\  
        \hline  
        $f$-sum & {-} & \multicolumn{1}{c|}{err=0.62\%} & \multicolumn{1}{c|}{err=-1.47\%} & \multicolumn{1}{c|}{err=-3.34\%} & {-} \\  
        \hline
        KK-sum & {-} & \multicolumn{1}{c|}{1.077 err = 7.74\%} & \multicolumn{1}{c|}{1.072 err=7.2\%} & \multicolumn{1}{c|}{0.956 err=-4.37\%} & {-} \\  
        \hline  
    \end{tabular}%
    }
    \caption{Comparison of the expected number of electrons in each excitation level of Cr, and the corresponding effective number of electrons $N_{\rm eff}$, as obtained  with the MELF-GOS model fit to the experimental optical ELF from Xu \textit{et al.} \cite{Xu2018}, Palik and Ghosh \cite{Palik1999} and Wehenkel and Gauthé \cite{Vehenkel1974}. The values of the $f$- and KK-sum rules appear in the last three rows.}
    \label{tab:comparison-Cr}
\end{table}

Table \ref{tab:comparison-Cr} gathers the total and partial effective numbers of electrons calculated for Cr, as well as the evaluation of the KK sum rule. The errors in the $f$-sum rule are around or less than $3$\% for all three sources of the ELF (which is acceptable), with the lowest error given by Xu \textit{et al.}'s data \cite{Xu2018}. The errors in the KK-sum rule are a bit larger, but always lower than $10$\%, with Wehenkel and Gauthé's ELF \cite{Vehenkel1974} yielding the lowest value.
The plasmon is most likely formed by electrons coming from the outermost $4s$ and $3d$ bands, containing respectively $1$ and $5$ electrons. Thus, these three excitations should contribute around $6$ effective electrons in total; all sources of the ELF underestimate this value, with Xu \textit{et al.}'s giving the closest number. 
As for the rest of the bands, effective numbers of electrons for all ELF are reasonably close to the expected number of electrons, except for underestimations in the innermost shells $2s$ and $1s$. It is well known that inner shells described by atomic GOS systematically underestimate the number of electrons due to Pauli’s exclusion principle. Since inner-shell electrons cannot be excited to occupied outer shells, they lose some amount of GOS, while the outer-shell electrons cannot fall into the occupied inner shells,
losing some amount of negative GOS \cite{Egerton2011}. As a consequence, the lack of inner-shell electrons must be distributed among the outer bands. These electrons seem to be distributed in the $3p$ band in the case of Xu \textit{et al.}'s ELF, and in the $3s$ band in the ELF from the other sources. 

\begin{figure}
	\centering
	\includegraphics[width=.56\textwidth]{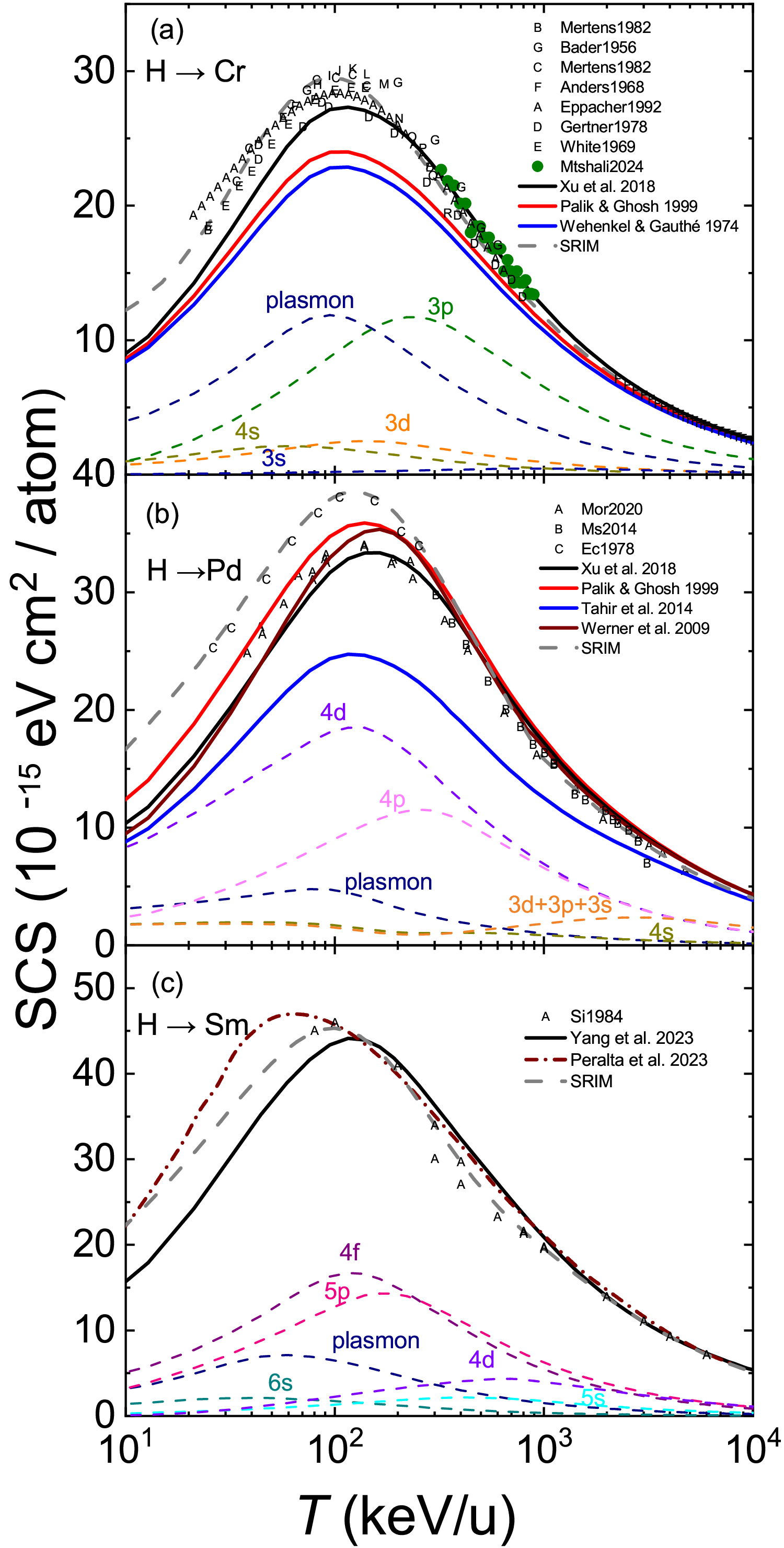}
	\caption{Comparison of experimental and calculated stopping cross sections of Cr (a), Pd (b), and Sm (c) for a proton beam, as a function of projectile energy $T$. Solid lines correspond to calculations from the ELF indicated in the labels. Dashed lines represent the contribution of each shell to the stopping cross section. Symbols represent available experimental data collected in \cite{PaulDatabase}.}
	\label{FIG:2}

\end{figure}

Figure \ref{FIG:2}(a) depicts by lines the Cr stopping cross sections (SCS, the stopping power divided by the atomic density of the target) for protons derived for the different optical ELF data. Dashed lines represent the contributions coming from the excitation of the different shells, according to the MELF-GOS fitting to the Xu \textit{et al.}'s ELF. Letter symbols correspond to the experimental stopping powers compiled in the late Helmut Paul's (currently IAEA) database \cite{PaulDatabase}, while green circles are the most recent experimental data by Mtshali \textit{et al.} \cite{Mtshali2024}. Calculations based on Palik and Ghosh's and Wehenkel and Gauthé's ELF clearly underestimate the experimental SCS, while Xu \textit{et al.}'s ELF gives results close to the IAEA's data and in excellent agreement with the most recent determinations. Our best results are close to SRIM values \cite{Ziegler2008} (gray dashed line in the figure) at high energies, even though SRIM results overestimate ours at the SCS maximum and below, in line with recent calculations by Peralta \textit{et al.} \cite{Peralta2022} (not shown to avoid confusion in the already crammed figure).

\begin{figure}
	\centering
	\includegraphics[width=0.7\textwidth]{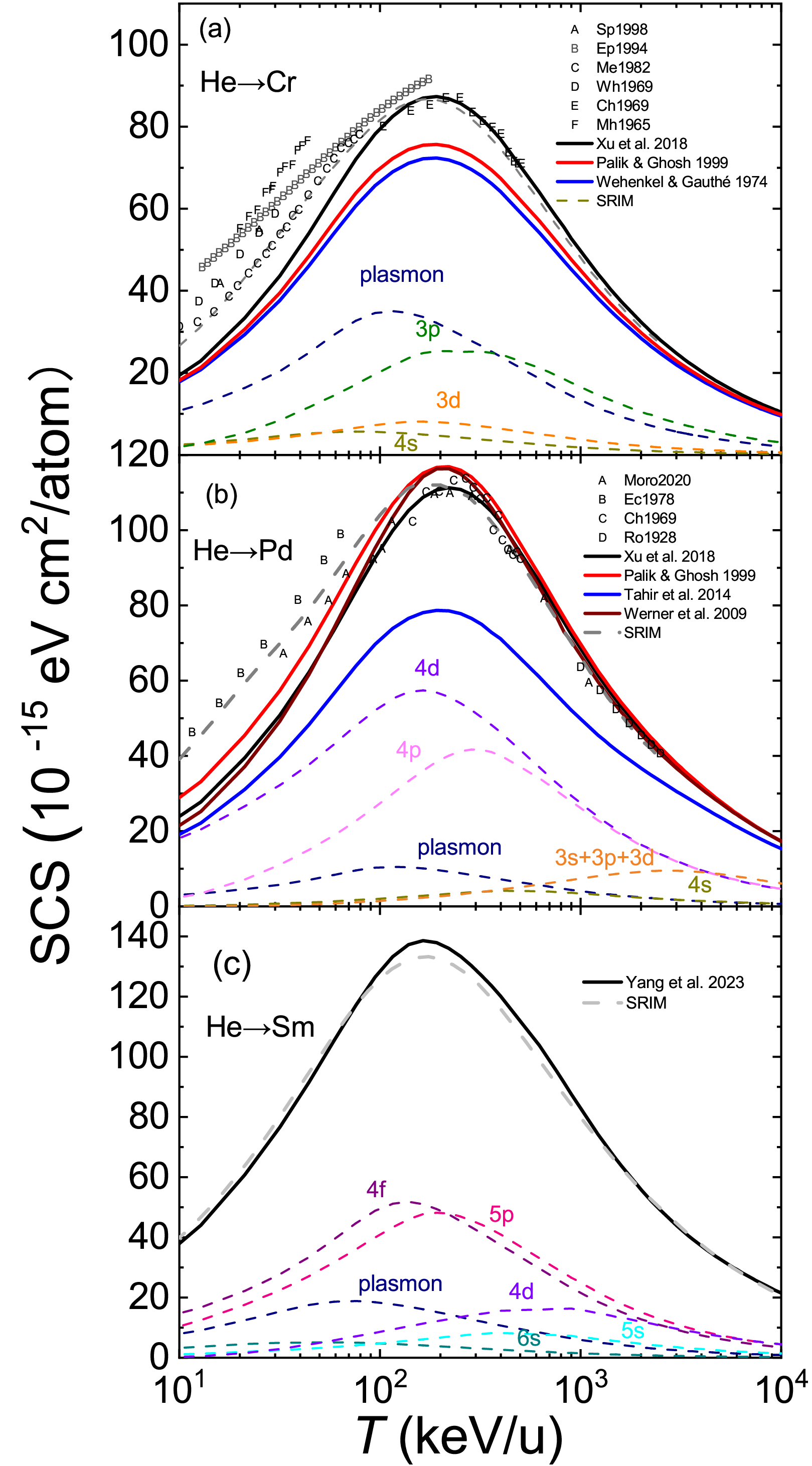}
	\caption{Comparison of experimental and calculated electronic stopping cross section of Cr (a), Pd (b), and Sm (c) for a He beam, as a function of projectile energy $T$. Solid lines correspond to calculations from the ELF indicated in the labels. Symbols represent experimental data from \cite{PaulDatabase}.}
	\label{FIG:3}
\end{figure}

Something similar happens with the SCS of Cr for He ions, shown in Fig. \ref{FIG:3}(a). Again, lines correspond to calculations using the different ELF data, while letter symbols depict the IAEA's compiled experimental data \cite{PaulDatabase}. As for protons, calculations based on Xu \textit{et al.}'s ELF give the best agreement with the experiments around the stopping maximum, while the ELF by Palik and Ghosh and Wehenkel and Gauthé clearly underestimate them. Our best results are similar to SRIM values \cite{Ziegler2008} (gray dashed line), except below the maximum.

\begin{figure}
	\centering
	\includegraphics[width=0.58\textwidth]{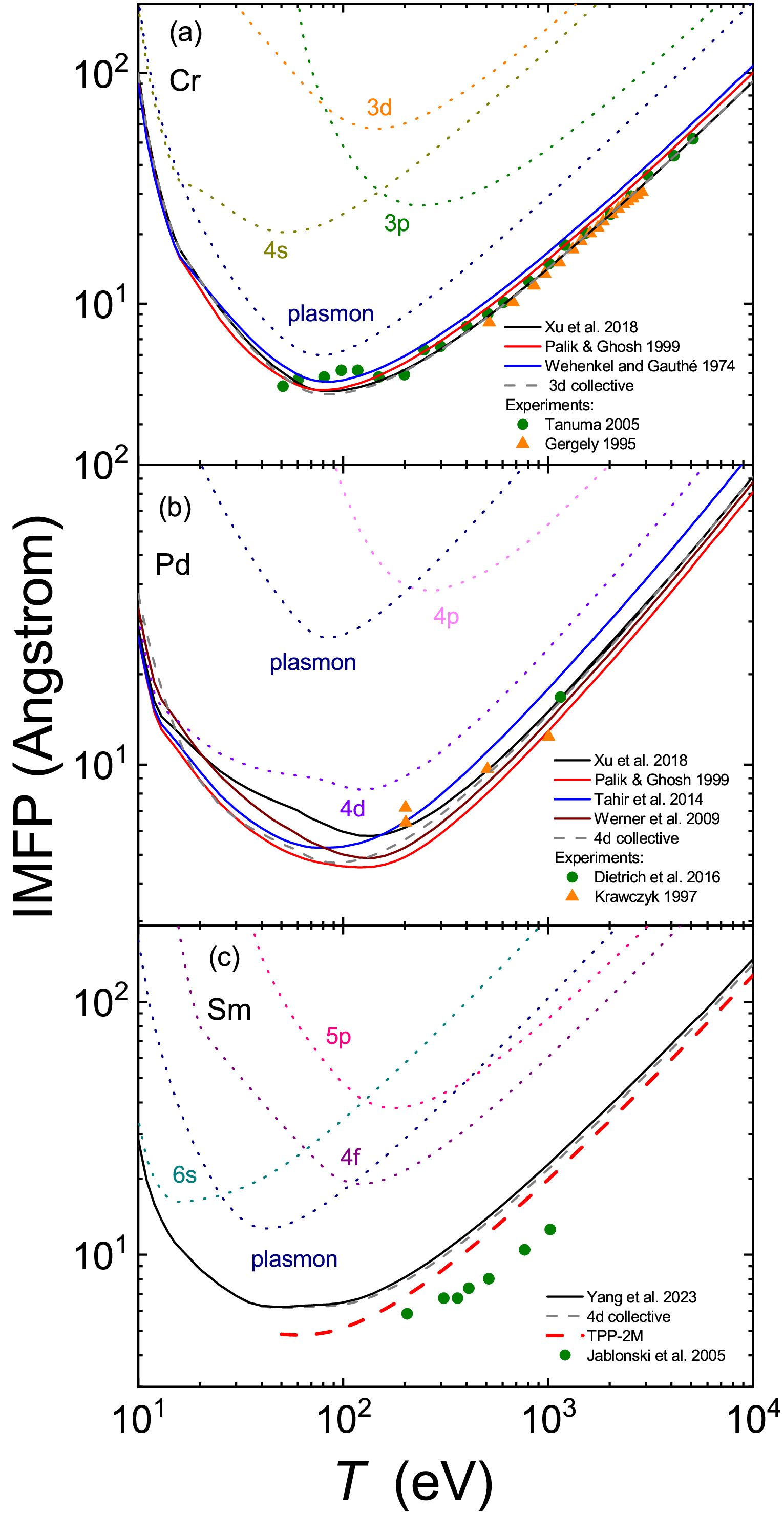}
	\caption{Electron IMFP in (a) Cr, (b) Pd and (c) Sm, as a function of the projectile energy $T$. The solid lines represent our calculations from different ELF, as indicated in the labels. The dashed line correspond to the predictive TPP-2M formula \cite{Tanuma1994}, and  symbols represent experimental results. The gray dashed line correspond to the $3d$ and $4d$ collective excitation for Cr and Pd, respectively, calculated with ELF of Xu et al. \cite{Xu2018} and the $4d$ for Sm calculated with Yang et al. \cite{Yang2023}. Dotted lines depict the contribution of each electronic shell to the inelastic mean free path. }
	\label{FIG:4}
\end{figure}

Finally, Fig. \ref{FIG:4}(a) shows by lines the calculated electron inelastic mean free path (IMFP) in Cr, compared to a collection of experimental measurements, depicted by symbols \cite{Gergely1995,Tanuma2005,Iakoubovskii2008}. The minimum of the mean free path is correctly reproduced by all three sets of optical ELF. Calculations based on Xu \textit{et al.}'s ELF match perfectly the high energy electron mean free path by Gergely \textit{et al.} \cite{Gergely1995} and Tanuma \textit{et al.} \cite{Tanuma2005}. All ELF provide a mean free path $\sim 1400$-$1450$ \AA \, at $200$ keV, which is of the same order of the experimental value reported by Iakoubovskii \textit{et al.} of $1040$ \AA \, \cite{Iakoubovskii2008} (the difference may be due to the disregard of relativistic effects in the current calculations). Even though Palik and Ghosh's data give place to a close result, both this ELF and, especially, Wehenkel and Gauthé's produce results slightly overestimating the experimental data at high energies. It should be noted that in Refs. \cite{deVera2019JPCC,deVera2023} it was discussed the possible consideration of the $3d$ electron excitation as of collective character. Accounting for this effect produces just a small change in the electron mean free path for Cr, which is shown in Fig. \ref{FIG:4}(a) as a green dashed line for the case of Xu \textit{et al.}'s ELF. Such a small difference does not have any impact in the preceding discussion.

Summarizing, both the Cr SCS around the maxima and the high energy electron mean free path are mainly influenced by the excitation of the plasmon and the $3p$ band. While the plasmon contribution to the ELF is rather similar for the three experimental sources (see Table \ref{tab:comparison-Cr}), Xu et al.'s ELF presents a larger contribution for the $3p$ band, which makes the calculated ions' SCS approach much more to the experimental data. The same happens with the high energy electron mean free path. Also, Xu \textit{et al.}'s ELF gives 
the lowest error in the total $f$-sum rule. All these facts together seem to justify the much better reproduction of SCS for protons and helium ions and of IMFP for electrons. It should be noted, in any case, the relevance of the excitation of the ``semi-valence'' $3p$ band: even though having a somewhat larger binding energy, this shell appreciably contributes to the valence-band ELF, giving place to large energy transfers with high frequency. As the stopping power integral, Eq. (\ref{eq:Sq}), involves the energy transfer $E$, the $3p$ band provides a large contribution to the SCS.

\subsection{Palladium}
\label{sect:Pd} 

The ELF of Pd is more complex (Fig. \ref{FIG:1}(b)), with a first feature around $10$ eV being most intense in Tahir \textit{et al.}'s TEELS data \cite{Tahir2014}.
Two intense peaks follow between $20$ and $40$ eV, now being Werner \textit{et al.}'s REELS ELF the most intense. After $40$ eV, Werner \textit{et al.}'s data present several intense peaks, while Xu \textit{et al.}'s RMC and Tahir \textit{et al.}'s ELF give a single broad feature with onset around $50$ eV, being more intense in the case of Xu \textit{et al.}'s. Except for the first peak, all other features are significantly less intense in the case of Tahir \textit{et al.}'s data as compared to the rest of sources.

The origin of the different excitations is assigned by assessing partial $f$-sum rules. The peak around $25$ eV is assumed to be due to the plasmon \cite{Egerton2011}, the rest of the structure below $50$ eV is assigned to the excitation of the $4d$ electrons, and the excitation with onset at this energy corresponds to the $4p$ electrons. Table \ref{tab:comparison-Pd} shows the total and partial effective numbers of electrons and the assessment of the KK-sum rule. All ELF sources fulfill the $f$-sum rule with errors below $1$\%, while the KK-sum rule errors are a bit larger, with Palik and Ghosh's data giving the largest one ($19$\%), Xu \textit{et al.}'s being moderate ($3.1$\%) and Tahir \textit{et al.}'s presenting a very low error ($0.72$\%). The outermost electrons are $4d$, which are the ones contributing to the plasmon. The sum of these two excitations, thus, should give around $10$ electrons. Palik and Ghosh's data 
and Xu \textit{et al.}'s ELF 
give the closest effective numbers of electrons. As for the ``semi-core'' $4p$ shell, Palik and Ghosh's ELF gives a very close number of electrons, Werner \textit{et al.}'s and Xu \textit{et al.}'s give a slight overestimation, while Tahir \textit{et al.}'s clearly underestimate the number of electrons in this excitation. It should be remembered that, in the case of Cr, the ``semi-core'' excitation contained an excess of electrons due to the underestimation in the inner shells as a result of the Pauli's exclusion principle. This may be the case here as well.

\begin{table}[h!]
    \centering
    \renewcommand{\arraystretch}{1.5}
    \setlength{\tabcolsep}{10pt}
    \resizebox{\textwidth}{!}{%
    \begin{tabular}{|c|c|c|c|c|c|c|}
        \hline
        Outer shell &  Expected electrons & Tahir \textit{et al.} \cite{Tahir2014} & Palik and Ghosh \cite{Palik1999} & Werner \textit{et al.} \cite{Werner2009} & Xu \textit{et al.} \cite{Xu2018} & Binding energy (eV) \\
        \hline \hline
        $4d$ & 10 & 4.03 & 8.83 & 4.98 & 7.96 & -\\
        \hline
        plasmon &  & 2.23 & 2.69 & 2.37 & 1.03 & - \\
         & & & & & & \\
        \hline
        $4p$ & 6 & 4.55 & 6.01 & 7.43 & 7.87 & 50.88 (for Xu is 53.06) \\
         & & & & & & \\
        \hline
        $4s$ & 2 & 4.61 & 1.10 & 1.64 & 1.44 & 87.07 \\
        \hline
        $3d$, $3p$, $3s$ & 18 & 23.82 & 20.85 & 23.066 & 21.15 & 335.23 \\
        \hline \hline
        Inner shell &   &  &  &  &  &  \\
        \hline \hline
        $2p$ &  6 & 4.1 & 4.1 & 4.1 & 4.1 & 3251.60 \\
        \hline
        $2s$ & 2 & 1.31 & 1.31 & 1.31 & 1.31 & 3603.96\\
        \hline
        $1s$ & 2 & 1.15 & 1.15 & 1.15 & 1.15 & 24350.23 \\
        \hline \hline
        Total & 46 & 45.76 & 46.03 & 46.03 & 46.00 &\\
        \hline  
        $f$-sum & {-} & \multicolumn{1}{c|}{err=$-0.51$\%} & \multicolumn{1}{c|}{err=$0.07$\%} & \multicolumn{1}{c|}{err=$0.07$\%} & \multicolumn{1}{c|}{err=0.02\%}& {-} \\  
        \hline
        KK-sum & {-} & \multicolumn{1}{c|}{1.007 err = 0.75\%} & \multicolumn{1}{c|}{1.19 err=19\%} & \multicolumn{1}{c|}{1.06 err=5.85\%} & \multicolumn{1}{c|}{1.03 err=3.1\%}& {-} \\ 
        \hline
    \end{tabular}%
    }
    \caption{Comparison of the expected number of electrons in each excitation level of Pd, and the corresponding effective number of electrons $N_{\rm eff}$, as obtained  with the MELF-GOS model fit to the experimental optical ELF from Tahir \textit{et al.} \cite{Tahir2014}, Palik and Ghosh \cite{Palik1999}, Werner \textit{et al.} \cite{Werner2009} and Xu \textit{et al.} \cite{Xu2018}. The values of the $f$- and KK-sum rules appear in the last three rows.}
    \label{tab:comparison-Pd}
\end{table}

Such an underestimation of the $4p$ shell in Tahir \textit{et al.}'s data has a dramatic impact on the SCS for both H and He projectiles, Figs. \ref{FIG:2}(b) and \ref{FIG:3}(b) respectively. While the rest of ELF sources provide SCS rather close to the cloud of experimental data points \cite{PaulDatabase}, Tahir \textit{et al.}'s ELF gives too low values, both around the maximum and even at intermediate ion energies. As for the rest of ELF sources, judging by the results both for H and He projectiles, Xu \textit{et al.}'s ELF again seems to provide the best agreement with the experimental information, particularly for He, even though in this case Palik and Ghosh's and Werner \textit{et al.}'s ELF give fair results. SRIM data (dashed gray lines) overestimate our best SCS calculation for H around and below the maximum, while for He the results are very similar, except below the maximum.

Figure \ref{FIG:4}(b) shows by lines the calculated electron IMFP in Pd, compared to a collection of experimental data, shown by symbols \cite{Krawczyk1997,Dietrich2016}. The minimum and the high energy mean free path from the ELF by Xu \textit{et al.} perfectly match the experiments by Krawczyk \textit{et al.} \cite{Krawczyk1997} and Dietrich \textit{et al.} \cite{Dietrich2016}. At $200$ keV, the experiment by Iakoubovskii \textit{ et al.} gives a value of $940$ \AA \, \cite{Iakoubovskii2008}, while Palik and Ghosh's and Xu \textit{et al.}'s ELF give calculated values between $1143$ and $1226$, and the other ELF a slight overestimation. The agreement between calculations based on Xu \textit{et al.}'s ELF and experiments by Krawczyk \textit{et al.} \cite{Krawczyk1997} is better when the excitation of the $4d$ electrons is not considered as a collective excitation (solid black line in the figure). In any case, the consideration of the $4d$ excitation as collective (green dashed line) \cite{deVera2019JPCC,deVera2023} does not change much the picture. Even though the scarcity of the experimental data (and the uncertainty related to the various data series from \cite{Krawczyk1997}) does not allow to draw strong conclusions on the quality of each ELF for Pd, this information should be considered together with that of H and He SCS (Figs. \ref{FIG:2}(b) and \ref{FIG:3}(b)). The ELF by Xu \textit{et al.} provides, in general, calculated results in almost perfect agreement with all experiments. While Palik and Ghosh's and Werner \textit{et al.}'s ELF give results relatively close to the latter, clearly the ELF by Tahir \textit{et al.} provides stopping quantities that are significantly underestimated.


In summary, the Pd SCS around the maxima and the high energy electron IMFP are mainly influenced by the excitation of the $4d$ and $4p$ bands. Xu \textit{et al.}'s ELF presents a close number of effective electrons in the $4d$ shell and plasmon excitation, while overestimating the number of electrons in the $4p$ shell. On the other hand, Tahir \textit{et al.}'s ELF lacks sufficient electrons in these two excitations. These facts seem to justify the good agreement of calculated stopping quantities based on Xu \textit{et al.}'s data (as well as on Palik and Ghosh's and Werner \textit{et al.}'s), while Tahir \textit{et al.}'s data provides energy loss quantities departing from experimental measurements.

\subsection{Samarium}
\label{sect:Sm} 

There is only one available source for the experimental optical ELF of Sm over a wide energy range, derived from the RMC method \cite{Yang2023}. Even though Palik and Ghosh's compilation includes some data \cite{Palik1999}, we do not make use of it, as it just covers experiments up to $\sim 3$ eV, complemented with atomic data above $30$ eV, missing the relevant intermediate part. The spectrum shown in Fig. \ref{FIG:1}(c) presents an intense plasmon peak around $15$ eV \cite{Egerton2011}, and the excitation of the $5p$ band has an onset around $20$ eV. The first peak below the plasmon has been assigned to the excitation of the $6s$ electrons, and the intense peak around $30$ eV to the excitation of the $4f$ band.

\begin{table}  
    \centering  
    \renewcommand{\arraystretch}{1.5}
    \setlength{\tabcolsep}{10pt}
    \begin{tabular}{|c|c|c|c|c|}
\hline  
Outer shell & Expected electrons & Yang \textit{et al.} \cite{Yang2023} & {Binding energy (eV)} \\  
\hline  \hline
$6s$ &  2 & 0.30 &  - \\  
\hline 
Plasmon & - & 1.34 & -  \\  
\hline 
$4f$ &  6 & 5.71 & 5.17 \\  
\hline 
$5p$ & 6 & 7.46 & 21.22 \\ 
\hline 
$5s$ & 2 & 3.02 & 37.28 \\  
\hline  
$4d$ & 10 & 11.22 & 128.98 \\ 
\hline 
$4p+4s$ & 8 & 6.19 & 247.34 \\  
\hline 
$3d+3p+3s$ & 18 & 21.27 & 1110.98 \\  
\hline \hline
Inner shell &  &  &  \\  
\hline  \hline
$2p$  & 6 & 3.21 & 7013.92  \\  
\hline  
$2s$  & 2 & 1.14 & 7736.90 \\  
\hline 
$1s$  & 2 & 1.04 & 46833.85 \\  
\hline \hline  
Total &  62 & 61.46 & - \\  
\hline  
$f$-sum & - & \multicolumn{1}{c|}{err=-0.88\%} & - \\  
\hline  
KK-sum & -  & \multicolumn{1}{c|}{0.99 err=-1.00\%} & - \\  
\hline  
\end{tabular} %

    \caption{Comparison of the expected number of electrons in each excitation level of Sm, and the corresponding effective number of electrons $N_{\rm eff}$, as obtained  with the MELF-GOS model fit to the experimental optical ELF from Yang \textit{et al.} \cite{Yang2023}. The values of the $f$- and KK-sum rules appear in the last three rows.}
    \label{tab:comparison-Sm}
\end{table}  

The $f$- and KK-sum rules are very well fulfilled (see Table \ref{tab:comparison-Sm}). The effective number of electrons comprised in the $6s$, plasmon and $4f$ excitations is very close to the expected number of $8$, while the effective numbers of electrons in the rest of excitations described by MELF are reasonable, some of them (e.g. $5p$) containing the excess of the electrons lacking from the inner-shells described by GOS, as it commonly happens due to Pauli's exclusion principle. 

For light ions' SCS, there only exist proton experimental data for this target \cite{PaulDatabase}, shown by letters in Fig. \ref{FIG:2}(c). The calculated SCS based on the RMC ELF, solid line, is rather close to this set of experimental measurements, reproducing well the high energy tail and the position of the maximum of the stopping power curve. It should be noted that current calculations are very close to the recent results by Peralta \textit{et al.} \cite{Peralta2023} (dash-dotted line in Fig. \ref{FIG:2}(c), based on the combination of the dielectric approach with a nonperturbative model for low energy projectiles), at least down to $\sim150$ keV/u, where we find the maximum. Below this energy, Peralta \textit{et al.} calculations still grow, predicting the maximum at $\sim60$ keV/u. Even though our results seem to agree with the trend of the experimental points, the scarcity of data prevents a more in-depth discussion. The SRIM values (gray dashed line) are in between of the previous calculations. For He projectiles (Fig. \ref{FIG:3}(c)), we did not find any experimental source or any other calculation, except for SRIM \cite{Ziegler2008}, which offers estimates very similar to our results in the entire energy range.

The available measurements of the electron IMFP of Sm \cite{Jablonski2005} are compared in Fig. \ref{FIG:4}(c) with our calculations, using the ELF provided by the RMC method \cite{Yang2023} (solid line), which is a little bit larger than the experimental values. However, these results are not very different from the predictions of the TPP-2M formula \cite{Tanuma1994}. It should be noted that the experiments are in the energy range from $200$ to $1000$ eV; at least above a few hundreds eV, the dielectric formalism should be accurate, and it has been shown to be reliable for several metals \cite{deVera2019JPCC,deVera2023}. Thus, there might be some degree of inaccuracy present in these IMFP measurements, particularly having into account the good agreement got for the proton SCS for Sm, as well as the electron IMFP shown above for Cr and Pd. 


In general, the calculated results for Sm seem satisfactory, even though there is a clear need of further experimental determinations of the optical ELF, as well as the ion stopping power and electron mean free path, in order to draw more definite conclusions.

\subsection{Mean excitation energies}
\label{sect:I}

The values of the mean excitation energy for each metal, obtained by means of Eq. (\ref{eq:I}) for the different available sets of ELF, are presented in Table \ref{tab:comparison-I} and compared with the reference values contained in the ICRU Report 37 \cite{ICRU37}. The expected differences in the $I$-values obtained from the different sources for each metal are rather small for Cr, with the result derived from Xu \textit{et al.} being closest to the ICRU Report 37 value. For Pd there is a large dispersion of $I$-values, with the result derived from Palik and Ghosh's ELF being the closest to ICRU Report 37, although the calculation based on Xu \textit{et al.}'s ELF is entirely compatible with the ICRU result, very close to the energies covered by the error bar range. However, the calculated $I$-value for Sm is in clear disagreement with that appearing in ICRU Report 37 (which is based on an interpolation from the elements with closest atomic numbers); despite the large difference, its consequence in the Bethe equation for the stopping power is practically inappreciable at high energies ($\geq 5$ MeV/u), although the SCS with the present $I=721.5$ eV is closer to the experimental results from $2$ to $5$ MeV//u.

\begin{table}  
    \centering  
    \renewcommand{\arraystretch}{1.5}
    \setlength{\tabcolsep}{10pt}
    \begin{tabular}{|c|c|c|c|}
\hline  
Spectrum source / Target & Cr & Pd & Sm \\  
\hline  \hline
Palik and Ghosh \cite{Palik1999} (OD) & 286.81 & 469.67 & - \\  
\hline 
Wehenkel and Gauthé \cite{Vehenkel1974} (TEELS) & 301.38 & - & - \\  
\hline  
Tahir et al. \cite{Tahir2014} (REELS) & - & 692.73 & - \\  
\hline 
Werner et al. \cite{Werner2009} (REELS) & - & 537.18 & - \\ 
\hline 
Xu et al. \cite{Xu2018} (RMC) & 255.64 & 507.31 & - \\  
\hline 
Yang et al. \cite{Yang2023} (RMC) & - & - & 721.53 \\ 
\hline
ICRU Report 37 \cite{ICRU37} & $257\pm{10}$ & $470\pm20$ & $574$ \\ 
\hline  
\end{tabular} %

    \caption{Mean excitation energy $I$ (in eV) of the metals discussed in this work, obtained by the MELF-GOS model from the different sources of the optical ELF. The last row coantains the $I$-value provided by ICRU Report 37 \cite{ICRU37}.}
    \label{tab:comparison-I}
\end{table}

\section{Summary and conclusions}
\label{sect:Conclusions}

As clearly seen from the previous results, the differences in intensity and width of the different electronic excitations of Cr, Pd, and Sm, contained in their optical ELF obtained from different experimental sources, give place to sizable differences in the calculated stopping quantities analyzed in this work for each metal (H and He ions SCS and electron IMFP). For Cr and Pd, the availability of experimental values for these energy loss quantities allows then the evaluation of the most accurate source for the optical ELF, while the RMC ELF-based results for Sm allow to confirm the good performance of this particular method. The correctness of each measured ELF could be also assessed by comparison with the \textit{ab initio} calculation methods currently available, which have shown high accuracy \cite{Pitarke2000,Botti2007,Yan2011,Taioli2021JPCL,Pedrielli2021,Taioli2023} and, for some materials, have been useful to identify the most accurate experimental measurement \cite{Pedrielli2021}. Unfortunately, we could not find in the literature first principles calculations of the ELF for Cr, Pd or Sm.

The consistency of each ELF has been assessed by means of partial and the total $f$-sum rule, as well as the KK-sum rule. Unfortunately, these integrals do not suffice to make a proper comparison of different experimental sources as, in general, total errors are always within a few percent for all the optical ELF. Partial sum rules are useful to assign the different features of the ELF to the excitation of electrons from particular bands. The partial effective number of electrons for each band are closer to the expected number of electrons for some sources than for others. Still, all sources give reasonable numbers, and this information alone neither suffices to judge for the most accurate source of data.

The observed differences in the features of the ELF for Cr and Pd give place to calculated stopping quantities which, for some sources agree very well with the different sets of experimental data while, for others, do not. Particularly, the ELF obtained from the reverse Monte Carlo method have always led to calculations in very good agreement with all stopping quantities measured experimentally for Cr, Pd, and Sm.
In a previous work \cite{deVera2023} we already observed that, for Fe, the reverse Monte Carlo ELF \cite{Xu2017} was the one best reproducing the experimental H stopping power too, which is consistent with present results. 
It should be noted that these results do not necessarily invalidate the ELF obtained from methods different from RMC; for example, the ELF for Pd from optical data by Palik and Ghosh and from REELS experiments by Werner \textit{et al.} give results very close to that from RMC, while the REELS derived data by Tahir \textit{et al.} clearly underestimate experimental data. In general, the calculation of the electron IMFP is less affected by the particular ELF used, but that of the SCS is critical. For Cr, the available optical data and TEELS ELF give H and He SCS that clearly underestimate the measurements.

We do not know the exact reason why the RMC method seems to work better than the others for obtaining the ELF of a metal. However, the characteristics of each experimental technique may justify the differences. It is well known that optical measurements are very sensitive to the sample surface roughness, as well as to potential surface contamination, as they are typically performed in ambient conditions. Having also into account that data acquisition in a wide energy range usually requires different measurements, possibly with different associated uncertainties, electron spectroscopy has become the method of choice for the ELF determination. Indeed, electron spectroscopy requires just one measurement to asses the ELF in a very wide energy range, it is not so sensitive to the surface roughness, and it is by default performed in ultra high vacuum conditions which avoid contamination. Still, the analysis of electron energy loss spectra requires some amount of modeling, particularly in the case of REELS, to account for multiple scattering effects and the influence of surface excitations. In this sense, the RMC method, using detailed Monte Carlo simulations to account for these effects, and using a self-consistent approach to arrive to the optimal ELF parameters which best reproduce the REELS spectrum \cite{Xu2017}, seems to work optimally for the electronic excitation spectrum determination. RMC has been successfully applied for several metals \cite{Xu2017,Xu2018,Li2023,Yang2023} as well as for oxides \cite{Da2013,Gong2024}, so the conclusions extracted from the present work for Cr, Pd, and Sm deserve further investigation for other materials.

Finally, it should be stressed the important role that ``semi-core'' excitations play in the energy loss quantities of heavy metals, particularly in the SCS. We refer to these excitations as the ones presenting significant binding energies (e.g. around $20$-$50$ eV), i.e. the $3p$ excitation in Cr, the $4p$ in Pd and the $5p$ in Sm. While the innermost shells typically have low intensities and contribute less to the stopping cross section, these ``semi-core'' excitations still present relatively large intensities in the $40$-$100$ eV range, giving a large contribution to the energy loss. Interestingly, for Cr and Pd, the successful reverse Monte Carlo ELF give place to an overestimation of the effective number of electrons in these excitations, while the optical data, presenting the largest underestimations in the stopping cross section, tend to yield effective numbers of electrons below the expected ones. It is worth to remember that it is common that the innermost shells present a lack of effective electrons due to the Pauli's exclusion principle, as discussed above. Therefore, it is normal for the missing electrons to be redistributed among the outer shells, so an excess of electrons in the ``semi-core'' excitations may be considered normal, and possibly even necessary if the ELF is to correctly predict the stopping cross sections, as apparent from present results.

The results and discussions presented in this work have shown the powerful capacity of the reverse Monte Carlo method for obtaining the ELF from REELS experiments. Thus, it is recommended to be used for the calculation of stopping quantities of metals when conflicting sources for the ELF are present, even though other methodologies for the ELF determination also provide fair results in some cases. For this purpose, we provide in Tables \ref{tab:params-Cr}, \ref{tab:params-Pd} and \ref{tab:params-Sm}, for Cr, Pd and Sm, respectively, the MELF-GOS fitting parameters to the respective reverse Monte Carlo ELF, so they can be used by other researchers.

\section*{Acknowledgements}

This work is part of the R\&D project no. PID2021-122866NB-I00 funded by the Spanish Ministerio de Ciencia e Innovación (MCIN/AEI/10.13039/501100011033/) and by the European Regional Development Fund (``ERDF A way
of making Europe''), as well as of the R\&D project no. 22081/PI/22 funded by the Autonomous Community of the Region of Murcia through the call ``Projects for the development of scientific and technical research by competitive groups'', included in the Regional Program for the Promotion of Scientific and Technical Research (Action Plan 2022) of the Fundación Séneca – Agencia de Ciencia y Tecnología de la Región de Murcia. F.C. thanks the China Scholarship Council (grant No. 202306340115) for financial support. We thank Dr. T. F. Yang (University of Science and Technology of China, Hefei, People’s Republic of China) for providing the data about the ELF of Sm contained in Ref. \cite{Yang2023}.



\begin{table}[h!]
    \centering
    \renewcommand{\arraystretch}{1.5}
    \setlength{\tabcolsep}{10pt}
    \resizebox{\textwidth}{!}{%
    \begin{tabular}{|c|c|c|c|c|c|c|c|}
        \hline
         $i$-th MELF & Shell & $\hbar\omega_i$ (eV) & $\gamma_i$ (eV)  & $A_i$ (eV$^2$) & $A_i'$ & $E_{{\rm th},i}$ (eV) & $\Delta_i$ (eV$^{-1}$)\\
        \hline 
        1 & 4s & 10.88 & 5.44 & 11.11 & 0.09 & - & -\\
        \hline
        2 & 4s & 16.33 & 10.88 & 37.02 & 0.14 & - & -\\
        \hline
        3 & Plasmon & 25.58 & 10.88 & 388.7 & 0.59 & - & -\\
        \hline
        4 & 3d & 33.47 & 19.05 & 88.85 & 0.08 & 8.71 & 1.47\\
        \hline
        5 & 3d & 45.44 & 8.16 & 222.12 & 0.11 & 8.71 & 1.47\\
        \hline
        6 & 3p & 53.06 & 32.65 & 236.92 & 0.08 & 46.26 & 1.10\\
        \hline
        7 & 3p & 57.14 & 32.65 & 481.25 & 0.15 & 46.26 & 1.10 \\
        \hline
        8 & 3p & 66.66 & 25.85 & 222.12 & 0.05 & 46.26 & 1.10\\
        \hline
        9 & 3p & 108.84 & 48.98 & 133.27 & 0.01 & 46.26 & 1.10\\
        \hline
        10 & 3s & 236.73 & 244.89 & 192.5 & 0.003 & 74.01 & 0.74\\
        \hline
    \end{tabular}%
    }
    \caption{Target: Cr. Fitting parameters ($\hbar\omega_i$, $\gamma_i$, $A_i$, $E_{{\rm th},i}$, $\Delta_i$) for the MELF-GOS fit to experimental data by Xu \textit{et al.} \cite{Xu2018}. For the exact meaning of each parameter, the reader is referred to the detailed description provided in Ref. \cite{deVera2023}.}
    \label{tab:params-Cr}
\end{table}

\begin{table}[h!]
    \centering
    \renewcommand{\arraystretch}{1.5}
    \setlength{\tabcolsep}{10pt}
    \resizebox{\textwidth}{!}{%
    \begin{tabular}{|c|c|c|c|c|c|c|c|}
        \hline
         $i$-th MELF & Shell & $\hbar\omega_i$ (eV) & $\gamma_i$ (eV)  & $A_i$ (eV$^2$) & $A_i'$ & $E_{{\rm th},i}$ (eV) & $\Delta_i$ (eV$^{-1}$)\\
        \hline 
        1 & 4d & 8.44 & 4.08 & 11.11 & 0.16 & - & - \\
        \hline
        2 & 4d & 9.77 & 8.16 & 1.48 & 0.02 & - & -\\
        \hline
        3 & 4d & 19.05 & 19.05 & 88.85 & 0.24 & - & -\\
        \hline
        4 & Plasmon & 26.94 & 8.16 & 96.25 & 0.13 & - & -\\
        \hline
        5 & 4d & 33.47 & 8.16 & 125.87 & 0.11 & - & -\\
        \hline
        6 & 4d & 43.54 & 25.85 & 88.85 & 0.05 & - & -\\
        \hline
        7 & 4d & 46.26 & 27.21 & 429.42 & 0.20 & - & - \\
        \hline
        8 & 4p & 54.42 & 27.21 & 340.58 & 0.12 & 53.06 & 1.47 \\
        \hline
        9 & 4p & 73.47 & 27.21 & 547.88 & 0.10 & 55.78 & 1.47\\
        \hline
        10 & 4s & 100.68 & 48.98 & 118.46 & 0.01 & 87.07 & 0.37\\
        \hline
        11 & 4s & 190.47 & 272.10 & 40.72 & 0.001 & 87.07 & 0.37\\
        \hline
        12 & 3d & 544.20 & 544.20 & 2110.09 & 0.01 & 335.23 & 0.74\\
        \hline
    \end{tabular}%
    }
    \caption{Target: Pd. Fitting parameters ($\hbar\omega_i$, $\gamma_i$, $A_i$, $E_{{\rm th},i}$, $\Delta_i$) for the MELF-GOS fit to experimental data by Xu \textit{et al.} \cite{Xu2018}. For the exact meaning of each parameter, the reader is referred to the detailed description provided in Ref. \cite{deVera2023}.}
    \label{tab:params-Pd}
\end{table}

\begin{table}[h!]
    \centering
    \renewcommand{\arraystretch}{1.5}
    \setlength{\tabcolsep}{10pt}
    \resizebox{\textwidth}{!}{%
    \begin{tabular}{|c|c|c|c|c|c|c|c|}
        \hline
        $i$-th MELF & Shell & $\hbar\omega_i$ (eV) & $\gamma_i$ (eV)  & $A_i$ (eV$^2$) & $A_i'$ & $E_{{\rm th},i}$ (eV) & $\Delta_i$ (eV$^{-1}$)\\
        \hline 
        1 & 6s & 3.10 & 2.90 & 1.70 & 0.18 & - & - \\
        \hline
        2 & 6s & 7.10 & 6.90 & 4.22 & 0.08 & - & -\\
        \hline
        3 & 6s & 10.50 & 7.80 & 6.59 & 0.06 & - & -\\
        \hline
        4 & Plasmon & 14.30 & 6.97 & 55.53 & 0.27 & - & -\\
        \hline
        5 & 4f & 19.59 & 10.88 & 22.21 & 0.06 & 5.17 & 1.29\\
        \hline
        6 & 4f & 23.40 & 8.16 & 20.73 & 0.04 & 5.17 & 1.29\\
        \hline
        7 & 4f & 27.50 & 5.90 & 8.88 & 0.01 & 5.17 & 1.29 \\
        \hline
        8 & 4f & 33.66 & 10.88 & 185.10 & 0.16 & 5.17 & 1.29\\
        \hline
        9 & 5p & 34.53 & 35.37 & 96.25 & 0.08 & 21.22 & 0.74\\
        \hline
        10 & 5p & 42.00 & 10.88 & 29.62 & 0.02 & 21.22 & 0.74\\
        \hline
        11 & 5p & 47.10 & 16.33 & 59.23 & 0.03 & 21.22 & 0.74\\
        \hline
        12 & 5p & 51.70 & 21.77 & 133.27 & 0.05 & 21.22 & 0.74\\
        \hline
        13 & 5s & 108.84 & 68.03 & 125.87 & 0.01 & 37.28 & 3.68\\
        \hline
        14 & 4d & 130.61 & 13.61 & 140.67 & 0.008 & 128.98 & 3.68\\
        \hline
        15 & 4d & 176.87 & 680.25 & 407.21 & 0.013 & 128.98 & 3.68\\
        \hline
        16 & 4p & 258.50 & 598.62 & 281.35 & 0.004 & 247.34 & 1.84\\
        \hline
        17 & 3d & 1170.03 & 897.93 & 1169.81 & 0.001 & 1110.98 & 0.55\\
        \hline
    \end{tabular}%
    }
    \caption{Target: Sm. Fitting parameters ($\hbar\omega_i$, $\gamma_i$, $A_i$, $E_{{\rm th},i}$, $\Delta_i$) for the MELF-GOS fit to experimental data by Yang \textit{et al.} \cite{Yang2023}. For the exact meaning of each parameter, the reader is referred to the detailed description provided in Ref. \cite{deVera2023}.}
    \label{tab:params-Sm}
\end{table}


\bibliography{library}

\end{document}